\documentclass{webofc}
\usepackage[varg]{txfonts}   
\usepackage{listings}
\usepackage{hyperref}
\begin{document}
\title{The $\gamma$ Dor stars as revealed by \emph{Kepler}: a key to reveal deep-layer rotation in A and F stars}
%
%

 \author{\firstname{S.J.A.J} \lastname{Salmon}\inst{1}\fnsep\thanks{\href{mailto:sebastien.salmon@ulg.ac.be}{\tt sebastien.salmon@ulg.ac.be}} 
\and
         \firstname{R.-M.} \lastname{Ouazzani}\inst{2}         
         \and
         \firstname{V.} \lastname{Antoci}\inst{2}
         \and
         \firstname{T.R.} \lastname{Bedding}\inst{3}         
         \and        
         \firstname{S.J.} \lastname{Murphy}\inst{3,2}}        
%
%

\institute{STAR Institute, Universit\'e de Li\`ege, All\'ee du Six Ao\^ut 19C, B-4000 Li\`ege, Belgium 
\and
Stellar Astrophysics Centre, Department of Physics and Astronomy, Aarhus University,\\
Ny Munkegade 120, DK-8000 Aarhus C, Denmark
           \and
           Sydney Institute for Astronomy (SIfA), School of Physics, University of Sydney,\\ Sydney, NSW 2006, Australia
          }

\abstract{%
The $\gamma$ Dor pulsating stars present high-order gravity modes, which make them important targets in the intermediate- and low-mass main-sequence region of the Hertzsprung-Russell diagram. Whilst we have only access to rotation in the envelope of the Sun, the g modes of $\gamma$ Dor stars can in principle deliver us constraints on the inner layers. With the puzzling discovery of unexpectedly low rotation rates in the core of red giants, the $\gamma$ Dor stars appear now as unique targets to explore internal angular momentum transport in the progenitors of red giants.
Yet, the $\gamma$ Dor pulsations remain hard to detect from the ground for their periods are close to 1 day. While the CoRoT space mission first revealed intriguing frequency spectra, the almost uninterrupted 4-year photometry from \emph{Kepler} mission eventually shed a new light on them. It revealed regularities in the spectra, expected to bear signature of physical processes, including rotation, in the shear layers close to the convective core. We present here the first results of our effort to derive exploitable seismic diagnosis for mid- to fast rotators among $\gamma$ Dor stars. We confirm their potential to explore the rotation history of this early phase of stellar evolution.    
}
\maketitle

\section{Introduction}\label{sec:intro}

In the last decade, asteroseismology has gathered in a wealth of results thanks, among others, to the CoRoT and \emph{Kepler} space missions. In particular, with the detection of pulsations in large samples of red giant stars, important progress has been made in our exploration of this more advanced phase of stellar evolution  (e.g. \cite{bedding11,beck11,mosser11,mosser12}).  The asteroseismic determination of the core rotation rate in red giants (\cite{beck12,mosser12bis,deheuvels15,mpdm16}) has revealed values significantly lower than theoretical predictions (e.g. \cite{tout11,sills11}). It has pinpointed the need of an efficient mechanism to extract angular momentum out of the core layers (\cite{eggenberger12,ceillier13,marques13,eggenberger17}). Several candidates, as e.g. angular momentum transport by internal gravity waves (\cite{fuller14}) or mixed modes (\cite{belk15}) were suggested but appear insufficient to explain the low constrast of rotation between core and surface observed in red giants. These new results push for exploring the preceding phases of the evolution to bring additional constraints on the unknown mechanism.


Among stars evolving to the giant branch, solar analogues would be the natural targets to explore the internal rotation during the main sequence. The inversion of the solar rotation profile is indeed one of the main achievement of helioseismology (e.g. \cite{brown89,koso97}). But the nature of its oscillations, acoustic modes, hampers the knowledge of the rotation profile near the core. Infering the internal rotation in other pulsating solar-like stars (\cite{benomar15,nielsen15}) deal with the same limitation. Here we explore the pulsations of sligthly more massive, A- and F-type, stars known as $\gamma$ Dor stars. Their gravity modes of pulsation propagate in the deepest layers of the star, and bear a signature of the rotation in these layers (\cite{bouabid13}). We present and summarize in the next section the first results obtained on the internal rotation in $\gamma$ Dor stars observed by \emph{Kepler}. We then conclude on how further investigation of these stars may pave the way to retrieve the evolution of rotation during the main sequence of intermediate- and low-mass stars.

\section{Revealing the internal rotation of $\gamma$ Dor stars}\label{sec:sec-1}

\begin{figure*}
\centering
\includegraphics{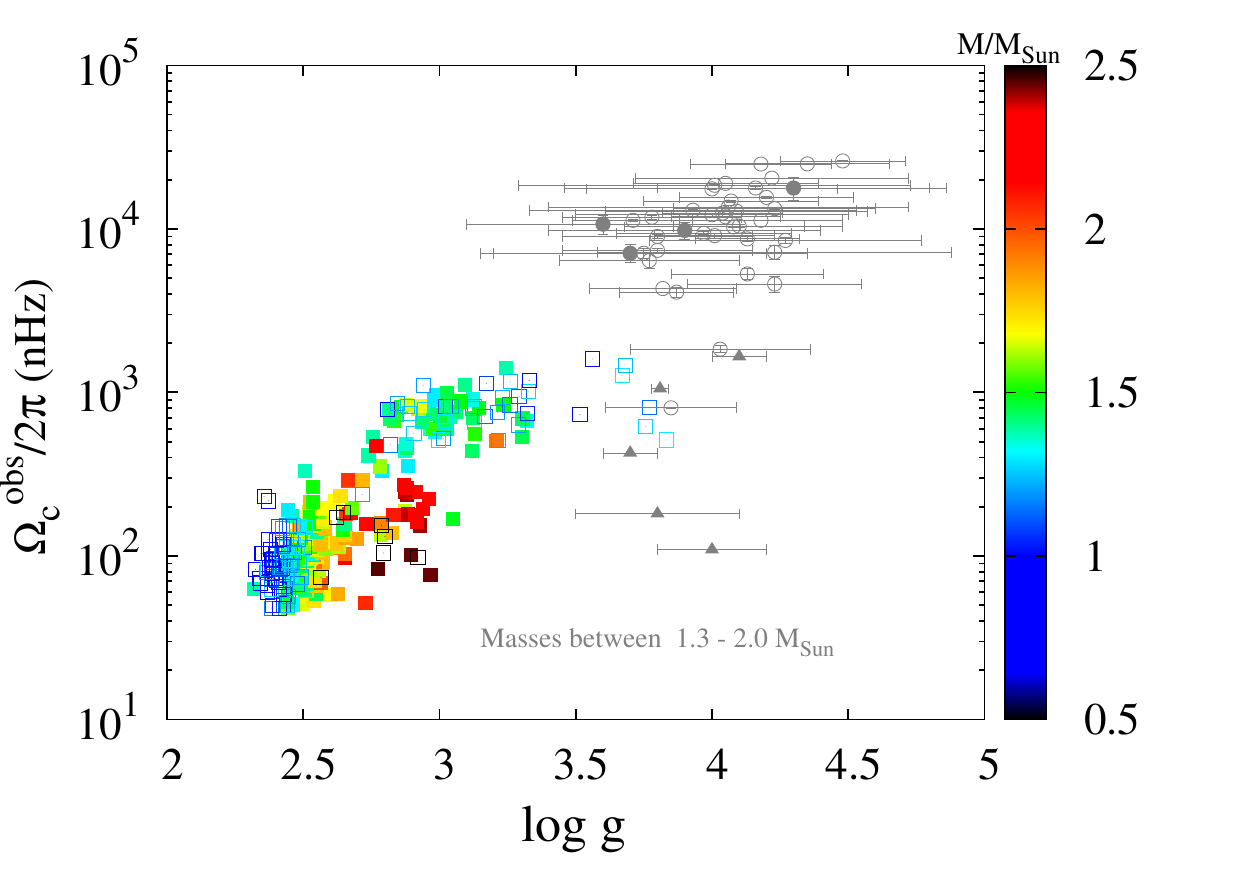}
\caption{Seismicly determined core rotation frequency as a function of the surface gravity of red giants and $\gamma$ Dor stars observed by \emph{Kepler}. Color-coded filled squares (1.3-2.5 M$_{\odot}$) and empty squares ($<$ 1.3 M$_{\odot}$) represent the subgiants and red giants analyzed in \cite{deheuvels14} and \cite{mosser12}. Grey triangles show the slowly rotating hybrid $\gamma$~Dor-$\delta$~Scuti stars from \cite{kurtz14,saio15,schmid15,keen15,murphy16}. Results from the $\gamma$ Dor studied by \cite{vanreeth16} and \cite{ouazzani17} are represented by grey open and filled circles, respectively. Masses of stars in grey are expected to be between $\sim$ 1.3 -- 2 M$_{\odot}$.}
\label{fig:fig-1}       
\end{figure*}

Remarkably, with the almost 4-years of uninterrupted data gathered by the \emph{Kepler} satellite, the resolution has now become sufficient to unravel the individual modes in the time-series analysis of $\gamma$ Dor stars. The first observational constraints on the internal rotation profile were obtained for $\gamma$ Dor-$\delta$ Scuti hybrid stars (\cite{kurtz14,saio15,schmid15,keen15,murphy16}. All of them are slow rotator with projected surface velocities of about 10 km/s. Thanks to the detection of rotational splitting in both their g ($\gamma$ Dor) and p mode ($\delta$ Scuti) regimes, these authors unveiled that they present all a nearly rigid rotation profile. The values of their deep-layer rotation frequencies are reported in Fig.\ref{fig:fig-1}, along with the core rotation from samples of subgiants (\cite{deheuvels14}) and red giants (\cite{mosser12}). This first sample of main-sequence objects present rather low values of rotation frequency in their deep layers  but show us it is possible to complete the picture of the rotation history from the main sequence to the helium-burning phase of intermediate-mass stars. 

For $\gamma$ Dor rotating faster, \cite{bouabid13} illustrated theoretically the importance of analysing the period spacing of g modes, on which rotation imprints its signatures under the form of regular patterns in the spacings. Recently, such features were indeed detected in the period spectra of \emph{Kepler} stars (\cite{bedding15,vanreeth15}). Motivated by these findings, we have developed in \cite{ouazzani17} practical tools to infer the rotation in the deep layers (close to the convective core) from the observed period spacings. In particular, we have demonstrated the existence of a relation between the slope of the linear trend in the period spacings and the rotation frequency, almost independent of the metallicity or the mass of the pulsators. 

Exploiting this relation resulted in the determination of the deep-layer rotation frequencies of 4 stars, represented in Fig.\ref{fig:fig-1}. Based on a similar approach, \cite{vanreeth16} determined rotation frequencies for another sample of \emph{Kepler} $\gamma$ Dor stars, also depicted in Fig.\ref{fig:fig-1}. Both studies reveals the existence of fast rotation in central layers of intermediate-mass stars, which was absent from the first results from the study of hybrid $\gamma$ Dor-$\delta$ Scuti stars.

From a first and qualitative comparison (P. Eggenberger, private communication), the values we found seem compatible with prediction from stellar models including rotation. The next step is now to refine the estimated rotation frequencies with detailed modelling of the observed stars, and proper comparison with theoretical predictions.

\section{Conclusion}\label{sec:conclusion}

With the unprecedent quality and length of the \emph{Kepler} data, we have now acces to the rotation of the deep layers in $\gamma$ Dor stars. These intermediate- and low-mass stars are fundamental to trace back the rotation history during the main-sequence. As a part to the puzzling question of the spin down of core rotation in red giants, we have access to an earlier phase of evolution to determine then what are the angular momentum transport mechanisms at work.  

To get firm constraints, we need better determination of the fundamental parameters, such as the surface gravity, of the $\gamma$ Dor stars observed by \emph{Kepler}. A better precision on these parameters is first necessary to evaluate their evolutionary stage on the main sequence. An ideal objective would be to build eventually the complete evolution of the rotation profile during the core hydrogen burning phase. Secondly, detailed modelling of the stars is requested to get precise profile of rotation with help of rotational kernels from a pulsation code including a complete treatment of the rotation effects (e.g. \cite{ouazzani12}). However, such a code requests an intense computing effort. Thus it is crucial to reduce the parameter space of the stellar models used for the modelling of observed pulsators.

Finally, to have a global and unbiased picture of the angular momentum transport mechanisms at work during the main sequence, we have to extend the sample of hybrid or $\gamma$ Dor stars from a thorough analysis of the \emph{Kepler} archives. We will then be able to make comprehensive comparison with stellar evolutionary codes and help determining whether their physical description of the rotational processes are adequate.


%




%
%

\end{document}